\pdfoutput=1

\documentclass{article}

\usepackage{arxiv}

\usepackage[utf8]{inputenc} 
\usepackage[T1]{fontenc}    
\usepackage{lmodern}        
\usepackage{hyperref}       
\usepackage{url}            
\usepackage{booktabs}       
\usepackage{amsfonts}       
\usepackage{nicefrac}       
\usepackage{microtype}      
\usepackage{graphicx}

\title{arulespy: Exploring Association Rules and Frequent Itemsets in
Python}

\author{
    Michael Hahsler
   \\
    Department of Computer Science \\
    Southern Methodist University \\
  Dallas, TX 75275 \\
  \texttt{\href{mailto:mhahsler@lyle.smu.edu}{\nolinkurl{mhahsler@lyle.smu.edu}}} \\
  }

\usepackage{color}
\usepackage{fancyvrb}

\DefineVerbatimEnvironment{Highlighting}{Verbatim}{commandchars=\\\{\}}
\newenvironment{Shaded}{}{}

\newcommand{\BuiltInTok}[1]{\textcolor[rgb]{0.00,0.50,0.00}{#1}}

\newcommand{\ControlFlowTok}[1]{\textcolor[rgb]{0.00,0.44,0.13}{\textbf{#1}}}

\newcommand{\DecValTok}[1]{\textcolor[rgb]{0.25,0.63,0.44}{#1}}

\newcommand{\ExtensionTok}[1]{#1}
\newcommand{\FloatTok}[1]{\textcolor[rgb]{0.25,0.63,0.44}{#1}}

\newcommand{\ImportTok}[1]{\textcolor[rgb]{0.00,0.50,0.00}{\textbf{#1}}}

\newcommand{\KeywordTok}[1]{\textcolor[rgb]{0.00,0.44,0.13}{\textbf{#1}}}
\newcommand{\NormalTok}[1]{#1}
\newcommand{\OperatorTok}[1]{\textcolor[rgb]{0.40,0.40,0.40}{#1}}

\newcommand{\StringTok}[1]{\textcolor[rgb]{0.25,0.44,0.63}{#1}}
\newcommand{\VariableTok}[1]{\textcolor[rgb]{0.10,0.09,0.49}{#1}}

\usepackage{longtable,booktabs,array}
\usepackage{calc} 
\usepackage{etoolbox}
\makeatletter
\patchcmd\longtable{\par}{\if@noskipsec\mbox{}\fi\par}{}{}
\makeatother
\IfFileExists{footnotehyper.sty}{\usepackage{footnotehyper}}{\usepackage{footnote}}
\makesavenoteenv{longtable}

\newlength{\cslhangindent}
\newlength{\csllabelwidth}
\newlength{\cslentryspacingunit} 
  {}%
  {\par}
\newenvironment{CSLReferences}[2] 
 {
  \setlength{\parindent}{0pt}
  \ifodd #1
  \let\oldpar\par
  \def\par{\hangindent=\cslhangindent\oldpar}
  \fi
  \setlength{\parskip}{#2\cslentryspacingunit}
 }%
 {}
\usepackage{calc}

\begin{document}
\maketitle

\begin{abstract}
The R package arules implements a comprehensive infrastructure for
representing, manipulating and analyzing transaction data and frequent
pattern mining. Is has been developed and maintained for the last 18
years and provides the most complete set of over 50 interest measures.
It contains optimized C/C++ code for mining and manipulating association
rules using sparse matrix representation. This document describes the
new Python package arulespy, which makes this infrastructure finally
available for Python users following Pythonic program writing style
including slicing and list comprehension.
\end{abstract}

\keywords{
    association rule mining
  }

\hypertarget{introduction}{%
\section{Introduction}\label{introduction}}

Association rule mining plays a vital role in discovering hidden
patterns and relationships within large transactional datasets.
Applications range from exploratory data analysis in marketing to
building rule-based classifiers. R (R Development Core Team 2005) users
have had access to the family of \texttt{arules} infrastructure packages
for association rule mining (Hahsler et al. 2011) for a long time. The
core packages are \texttt{arules} (Hahsler, Grün, and Hornik 2005) which
provides the infrastructure for representing, manipulating and analyzing
transaction data and frequent patterns (itemsets and association rules),
and \texttt{arulesViz} (Hahsler 2017), providing various visualization
techniques for association rules and itemsets. The packages are built on
contributed C code, like the implementaiton of the APRIORI algorithm and
the ECLAT algorithm provided by Christian Borgelt (Borgelt 2003), C++
sparse matrix code provided by the R \texttt{Matrix} package (Bates,
Maechler, and Jagan 2022), and custom C/C++ code implemented by the
\texttt{arules} team, and R interface code. For portability, all C and
C++ code has been updated to the latest standard (C++20 and C17). The
Python interface is based on \texttt{rpy2} (Gautier 2022). Much care has
been taken to translate R's functional interface into a Pythonic package
providing Python programmers with expected behavior.

With many data scientists needing to work with R, R markdown, but also
with Python and Jupyter notebooks, \texttt{arulespy} provides a native
and easy to install Python interface to the wide range of functionalists
provided by the R packages \texttt{arules} and \texttt{arulesViz}.

Several popular Python packages provide frequent pattern mining in
Python including in the popular \texttt{mlxtend} package (Raschka 2018),
but \texttt{arules} still provides a higher level of functionality in
terms of visualization options, and available infrastructure.

\hypertarget{package-installation}{%
\section{Package Installation}\label{package-installation}}

\texttt{arulespy} is based on the Python package \texttt{rpy2}, which
requires an R installation. \texttt{arulespy} is easily installed from
the Python Package Index using pip:

\begin{Shaded}
\begin{Highlighting}[]
\ExtensionTok{pip}\NormalTok{ install arulespy}
\end{Highlighting}
\end{Shaded}

This installation will take care of installing the needed R packages.
Note that the R packages are installed during the first time arulespy is
imported. This installation may require some time. Detailed installation
instructions can be found on the package's PyPI page (Hahsler 2023).

\hypertarget{overview-of-features}{%
\section{Overview of features}\label{overview-of-features}}

\hypertarget{pythonic-high-level-interface}{%
\subsection{Pythonic high-level
interface}\label{pythonic-high-level-interface}}

\texttt{arulespy} provides the computational infrastructure to represent
all data structures necessary to mine association rules. Agrawal,
Imielinski, and Swami (1993) introduced the problem of mining
association rules from transaction data as follows (the definition is
taken from Hahsler, Grün, and Hornik (2005)):

Let \(I = \{i_1,i_2,...,i_n\}\) be a set of \(n\) binary attributes
called items. Let \(D = \{t_1,t_2,...,t_m\}\) be a set of transactions
called the database. Each transaction in \(D\) has a unique transaction
ID and contains a subset of the items in \(I\). A rule is defined as an
implication of the form \(X \Rightarrow Y\) where \(X,Y \subseteq I\)
and \(X \cup Y = \emptyset\) are called itemsets. On itemsets and rules
several quality measures can be defined. The most important measures are
support and confidence. The support \(supp(X)\) of an itemset \(X\) is
defined as the proportion of transactions in the data set which contain
the itemset. Itemsets with a support which surpasses a user-defined
threshold \(\sigma\) are called frequent itemsets. The confidence of a
rule is defined as \(conf(X \Rightarrow Y) = supp(X \cup Y)/supp(X)\).
Association rules are rules with \(supp(X \cup Y) \ge \sigma\) and
\(conf(X) \ge \delta\) where \(\sigma\) and \(\delta\) are user-defined
thresholds. More measures to judge the quality or interestingness of
rules and itemsets have been described in the literature (see Tan,
Kumar, and Srivastava (2004), Geng and Hamilton (2006), Lenca et al.
(2007)). A complete list of available interest measures in
\texttt{arulespy} can be found in Hahsler (2005).

\texttt{arulespy} provides a high-level interface to transactions,
itemsets, and rules based on sparse matrices representing sets of
itemsets as with the class \texttt{itemMatrix}. Figure 1 shows the
implemented classes divided by module.

\begin{figure}

{\centering \includegraphics[width=0.9\linewidth]{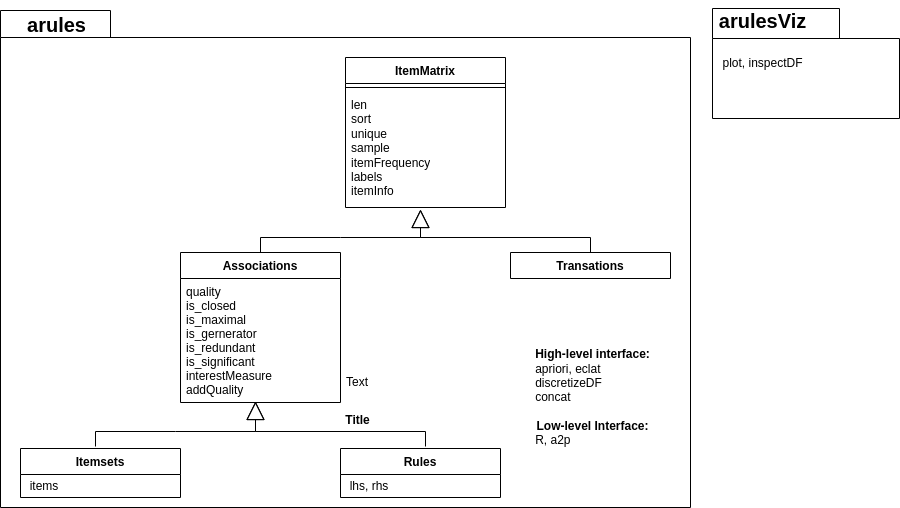} 

}

\caption{arulespy modules with Python classes.}\label{fig:classes}
\end{figure}

All classes provide Pythonic slicing using \texttt{{[}{]}}, ranges,
\texttt{len()}, methods for \texttt{sort()}, \texttt{unique()},
\texttt{sample()}, and methods to convert the data into different Python
data structures are provided. These include: \texttt{as\_df()} (a pandas
dataframe), \texttt{as\_matrix()} (a numpy matrix), \texttt{as\_dict()}
(a Python dictionary), and \texttt{as\_list()}. Associations in addition
provide methods to extract \texttt{quality()} information,
\texttt{items()} (\texttt{lhs()} and \texttt{rhs()} for rules), and to
determine if the association \texttt{is\_closed()},
\texttt{is\_maximal()}, \texttt{is\_generator()},
\texttt{is\_redundant()}, or \texttt{is\_significant()}.

Combining objects is modeled after \texttt{pandas.concat()} taking a
list of objects to combine.

For visualization, module \texttt{arulesViz} contains a \texttt{plot()}
function to produce visualizations via \texttt{ggplot} and interactive
HTML widgets to inspect rules (\texttt{inspectDF()}).

The package uses docstrings and Python help can be obtained using
\texttt{help()}.

\hypertarget{creating-transaction-data}{%
\subsection{Creating transaction data}\label{creating-transaction-data}}

To prepare transaction data, arulespy provides \texttt{discretizeDF()}
to prepare a pandas dataframe with numeric attributes by discretization
and \texttt{Transactions.from\_df()} to convert pandas dataframes into
sparse transaction representation.

\begin{Shaded}
\begin{Highlighting}[]
\ImportTok{from}\NormalTok{ arulespy.arules }\ImportTok{import}\NormalTok{ Transactions, apriori, parameters}
\end{Highlighting}
\end{Shaded}

The data need to be prepared as a pandas dataframe. Here we import the
Zoo dataset (Asuncion and Newman 2007) which contains the features of
101 animales as binary attributes, a numeric attribute for the number of
legs and nominal animal type. We show some of the attributes for the
first three animals.

\begin{Shaded}
\begin{Highlighting}[]
\ImportTok{import}\NormalTok{ pandas }\ImportTok{as}\NormalTok{ pd}

\NormalTok{df }\OperatorTok{=}\NormalTok{ pd.read\_csv(}\StringTok{\textquotesingle{}https://mhahsler.github.io/arulespy/examples/Zoo.csv\textquotesingle{}}\NormalTok{)}
\NormalTok{df[}\DecValTok{0}\NormalTok{:}\DecValTok{3}\NormalTok{]}
\end{Highlighting}
\end{Shaded}

\begin{longtable}[]{@{}
  >{\raggedleft\arraybackslash}p{(\columnwidth - 18\tabcolsep) * \real{0.0543}}
  >{\raggedright\arraybackslash}p{(\columnwidth - 18\tabcolsep) * \real{0.0870}}
  >{\raggedright\arraybackslash}p{(\columnwidth - 18\tabcolsep) * \real{0.1304}}
  >{\raggedright\arraybackslash}p{(\columnwidth - 18\tabcolsep) * \real{0.0870}}
  >{\centering\arraybackslash}p{(\columnwidth - 18\tabcolsep) * \real{0.0543}}
  >{\raggedleft\arraybackslash}p{(\columnwidth - 18\tabcolsep) * \real{0.0870}}
  >{\raggedright\arraybackslash}p{(\columnwidth - 18\tabcolsep) * \real{0.0870}}
  >{\raggedright\arraybackslash}p{(\columnwidth - 18\tabcolsep) * \real{0.1304}}
  >{\raggedright\arraybackslash}p{(\columnwidth - 18\tabcolsep) * \real{0.1196}}
  >{\raggedright\arraybackslash}p{(\columnwidth - 18\tabcolsep) * \real{0.1630}}@{}}
\toprule()
\begin{minipage}[b]{\linewidth}\raggedleft
\end{minipage} & \begin{minipage}[b]{\linewidth}\raggedright
hair
\end{minipage} & \begin{minipage}[b]{\linewidth}\raggedright
feathers
\end{minipage} & \begin{minipage}[b]{\linewidth}\raggedright
eggs
\end{minipage} & \begin{minipage}[b]{\linewidth}\centering
\ldots{}
\end{minipage} & \begin{minipage}[b]{\linewidth}\raggedleft
legs
\end{minipage} & \begin{minipage}[b]{\linewidth}\raggedright
tail
\end{minipage} & \begin{minipage}[b]{\linewidth}\raggedright
domestic
\end{minipage} & \begin{minipage}[b]{\linewidth}\raggedright
catsize
\end{minipage} & \begin{minipage}[b]{\linewidth}\raggedright
type
\end{minipage} \\
\midrule()
\endhead
0 & True & False & False & \ldots{} & 4 & False & False & True &
mammal \\
1 & True & False & False & \ldots{} & 4 & True & False & True &
mammal \\
2 & False & False & True & \ldots{} & 0 & True & False & False & fish \\
\bottomrule()
\end{longtable}

Next, we convert the animals to transactions. In this process, binary
attributes are converted into items, numeric attributes are discretized
into range items and nominal attributes are automatically
one-hot-encoded.

\begin{Shaded}
\begin{Highlighting}[]
\NormalTok{trans }\OperatorTok{=}\NormalTok{ Transactions.from\_df()}
\BuiltInTok{print}\NormalTok{(trans)}
\end{Highlighting}
\end{Shaded}

\begin{verbatim}
##    transactions in sparse format with
##     101 transactions (rows) and
##     25 items (columns)
\end{verbatim}

\begin{Shaded}
\begin{Highlighting}[]
\NormalTok{trans.as\_df()}
\end{Highlighting}
\end{Shaded}

\begin{longtable}[]{@{}
  >{\raggedleft\arraybackslash}p{(\columnwidth - 4\tabcolsep) * \real{0.0476}}
  >{\raggedright\arraybackslash}p{(\columnwidth - 4\tabcolsep) * \real{0.7905}}
  >{\raggedleft\arraybackslash}p{(\columnwidth - 4\tabcolsep) * \real{0.1619}}@{}}
\toprule()
\begin{minipage}[b]{\linewidth}\raggedleft
\end{minipage} & \begin{minipage}[b]{\linewidth}\raggedright
items
\end{minipage} & \begin{minipage}[b]{\linewidth}\raggedleft
transactionID
\end{minipage} \\
\midrule()
\endhead
1 &
\{hair,milk,predator,toothed,\ldots,legs={[}4,8{]},catsize,type=mammal\}
& 0 \\
2 & \{hair,milk,toothed,\ldots,legs={[}4,8{]},tail,catsize,type=mammal\}
& 1 \\
3 & \{eggs,aquatic,predator,\ldots,fins,legs={[}0,2),tail,type=fish\} &
2 \\
\ldots{} & \ldots{} & \ldots{} \\
\bottomrule()
\end{longtable}

For space reasons, we only show the first three transactions and omit
some of the items. Transactions are stored as a sparse matrix. The item
definitions show that the number of legs was discretized into 3 ranges
and the animal type was converted into 7 binary items.

\begin{Shaded}
\begin{Highlighting}[]
\NormalTok{trans.itemInfo()}
\end{Highlighting}
\end{Shaded}

\begin{longtable}[]{@{}rlll@{}}
\toprule()
& labels & variables & levels \\
\midrule()
\endhead
1 & hair & hair & TRUE \\
2 & feathers & feathers & TRUE \\
3 & eggs & eggs & TRUE \\
4 & milk & milk & TRUE \\
5 & airborne & airborne & TRUE \\
6 & aquatic & aquatic & TRUE \\
7 & predator & predator & TRUE \\
8 & toothed & toothed & TRUE \\
9 & backbone & backbone & TRUE \\
10 & breathes & breathes & TRUE \\
11 & venomous & venomous & TRUE \\
12 & fins & fins & TRUE \\
13 & legs={[}0,2) & legs & {[}0,2) \\
14 & legs={[}2,4) & legs & {[}2,4) \\
15 & legs={[}4,8{]} & legs & {[}4,8{]} \\
16 & tail & tail & TRUE \\
17 & domestic & domestic & TRUE \\
18 & catsize & catsize & TRUE \\
19 & type=amphibian & type & amphibian \\
20 & type=bird & type & bird \\
21 & type=fish & type & fish \\
22 & type=insect & type & insect \\
23 & type=mammal & type & mammal \\
24 & type=mollusc.et.al & type & mollusc.et.al \\
25 & type=reptile & type & reptile \\
\bottomrule()
\end{longtable}

We can easily slice transactions, sample form transactions, combine them
and find unique transactions using methods for the Python Transactions
class.

\hypertarget{mining-association-rules}{%
\subsection{Mining Association Rules}\label{mining-association-rules}}

The mining functions \texttt{apriori()} and \texttt{eclat()} are part of
the high-level interface. \texttt{apriori()} calls the APRIORI algorithm
implemented in the R package arules and performs all necessary
conversions. Parameters for the algorithm are specified as a Python
\texttt{dict} inside the \texttt{parameter()} function.

\begin{Shaded}
\begin{Highlighting}[]
\NormalTok{rules }\OperatorTok{=}\NormalTok{ apriori(trans,}
\NormalTok{                    parameter }\OperatorTok{=}\NormalTok{ parameters(\{}\StringTok{"supp"}\NormalTok{: }\FloatTok{0.01}\NormalTok{, }\StringTok{"conf"}\NormalTok{: }\FloatTok{0.7}\NormalTok{\}), }
\NormalTok{                    control }\OperatorTok{=}\NormalTok{ parameters(\{}\StringTok{"verbose"}\NormalTok{: }\VariableTok{False}\NormalTok{\}))  }

\BuiltInTok{print}\NormalTok{(rules)}
\end{Highlighting}
\end{Shaded}

\begin{verbatim}
##  set of 30438 rules
\end{verbatim}

We can inspect the top three confidence rules.

\begin{Shaded}
\begin{Highlighting}[]
\NormalTok{rules.sort(by }\OperatorTok{=} \StringTok{\textquotesingle{}confidence\textquotesingle{}}\NormalTok{)[}\DecValTok{0}\NormalTok{:}\DecValTok{3}\NormalTok{].as\_df()}
\end{Highlighting}
\end{Shaded}

\begin{longtable}[]{@{}
  >{\raggedleft\arraybackslash}p{(\columnwidth - 14\tabcolsep) * \real{0.0526}}
  >{\raggedright\arraybackslash}p{(\columnwidth - 14\tabcolsep) * \real{0.2368}}
  >{\raggedright\arraybackslash}p{(\columnwidth - 14\tabcolsep) * \real{0.1842}}
  >{\raggedleft\arraybackslash}p{(\columnwidth - 14\tabcolsep) * \real{0.1053}}
  >{\raggedleft\arraybackslash}p{(\columnwidth - 14\tabcolsep) * \real{0.1447}}
  >{\raggedleft\arraybackslash}p{(\columnwidth - 14\tabcolsep) * \real{0.1184}}
  >{\raggedleft\arraybackslash}p{(\columnwidth - 14\tabcolsep) * \real{0.0789}}
  >{\raggedleft\arraybackslash}p{(\columnwidth - 14\tabcolsep) * \real{0.0789}}@{}}
\toprule()
\begin{minipage}[b]{\linewidth}\raggedleft
\end{minipage} & \begin{minipage}[b]{\linewidth}\raggedright
LHS
\end{minipage} & \begin{minipage}[b]{\linewidth}\raggedright
RHS
\end{minipage} & \begin{minipage}[b]{\linewidth}\raggedleft
support
\end{minipage} & \begin{minipage}[b]{\linewidth}\raggedleft
confidence
\end{minipage} & \begin{minipage}[b]{\linewidth}\raggedleft
coverage
\end{minipage} & \begin{minipage}[b]{\linewidth}\raggedleft
lift
\end{minipage} & \begin{minipage}[b]{\linewidth}\raggedleft
count
\end{minipage} \\
\midrule()
\endhead
4 & \{type=amphibian\} & \{aquatic\} & 0.04 & 1 & 0.04 & 2.81 & 4 \\
5 & \{type=amphibian\} & \{legs={[}4,8{]}\} & 0.04 & 1 & 0.04 & 1.98 &
4 \\
6 & \{type=amphibian\} & \{eggs\} & 0.04 & 1 & 0.04 & 1.71 & 4 \\
\bottomrule()
\end{longtable}

The set of rules is rather large with a length of

\begin{Shaded}
\begin{Highlighting}[]
\BuiltInTok{len}\NormalTok{(rules)}
\end{Highlighting}
\end{Shaded}

\begin{verbatim}
##  30438
\end{verbatim}

Rules can be tested for many properties. For example, a rule is
improvement-based redundant if a more general rule with the same or a
higher confidence exists in the set (Bayardo, Agrawal, and Gunopulos
2000). The following code filters all redundant rules using Python list
comprehension reducing the set to about a third.

\begin{Shaded}
\begin{Highlighting}[]
\NormalTok{non\_redundant\_rules }\OperatorTok{=}\NormalTok{ rules[[}\KeywordTok{not}\NormalTok{ x }\ControlFlowTok{for}\NormalTok{ x }\KeywordTok{in}\NormalTok{ rules.is\_redundant()]]}
\NormalTok{non\_redundant\_rules.as\_df()}
\end{Highlighting}
\end{Shaded}

\begin{longtable}[]{@{}
  >{\raggedleft\arraybackslash}p{(\columnwidth - 14\tabcolsep) * \real{0.0588}}
  >{\raggedright\arraybackslash}p{(\columnwidth - 14\tabcolsep) * \real{0.3361}}
  >{\raggedright\arraybackslash}p{(\columnwidth - 14\tabcolsep) * \real{0.1513}}
  >{\raggedleft\arraybackslash}p{(\columnwidth - 14\tabcolsep) * \real{0.0924}}
  >{\raggedleft\arraybackslash}p{(\columnwidth - 14\tabcolsep) * \real{0.1176}}
  >{\raggedleft\arraybackslash}p{(\columnwidth - 14\tabcolsep) * \real{0.1008}}
  >{\raggedleft\arraybackslash}p{(\columnwidth - 14\tabcolsep) * \real{0.0672}}
  >{\raggedleft\arraybackslash}p{(\columnwidth - 14\tabcolsep) * \real{0.0756}}@{}}
\toprule()
\begin{minipage}[b]{\linewidth}\raggedleft
\end{minipage} & \begin{minipage}[b]{\linewidth}\raggedright
LHS
\end{minipage} & \begin{minipage}[b]{\linewidth}\raggedright
RHS
\end{minipage} & \begin{minipage}[b]{\linewidth}\raggedleft
support
\end{minipage} & \begin{minipage}[b]{\linewidth}\raggedleft
confidence
\end{minipage} & \begin{minipage}[b]{\linewidth}\raggedleft
coverage
\end{minipage} & \begin{minipage}[b]{\linewidth}\raggedleft
lift
\end{minipage} & \begin{minipage}[b]{\linewidth}\raggedleft
count
\end{minipage} \\
\midrule()
\endhead
1 & \{\} & \{tail\} & 0.74 & 0.74 & 1 & 1 & 75 \\
2 & \{\} & \{breathes\} & 0.79 & 0.79 & 1 & 1 & 80 \\
3 & \{\} & \{backbone\} & 0.82 & 0.82 & 1 & 1 & \\
\ldots{} & \ldots{} & \ldots{} & \ldots{} & \ldots{} & \ldots{} & & \\
11676 & \{predator,toothed, breathes,tail\} & \{catsize\} & 0.15 & 0.71
& 0.21 & 1.64 & 15 \\
11691 & \{eggs,toothed, breathes,tail\} & \{predator\} & 0.04 & 1 & 0.04
& 1.8 & 4 \\
\bottomrule()
\end{longtable}

\hypertarget{visualization}{%
\subsection{Visualization}\label{visualization}}

The set of rules is still relatively large, but visualization can help
analyzing the rules. The \texttt{arulespy} module \texttt{arulesViz} can
produce a wide range of visualizations for association rules (Hofmann
and Wilhelm 2001). It exports a plot function which produces
\texttt{ggplot2} plots (Wickham 2016).

\begin{Shaded}
\begin{Highlighting}[]
\ImportTok{from}\NormalTok{ arulespy.arulesViz }\ImportTok{import}\NormalTok{ plot, inspectDF}
\ImportTok{from}\NormalTok{ rpy2.ipython.ggplot }\ImportTok{import}\NormalTok{ image\_png}
\end{Highlighting}
\end{Shaded}

The standard plot is a scatter plot of rules using support and
consequence on the axes and lift for color shading. \texttt{ggplot}
objects can be rendered directly in a Jupyter notebook code cell using
\texttt{image\_png()}.

\begin{Shaded}
\begin{Highlighting}[]
\NormalTok{gg }\OperatorTok{=}\NormalTok{ arulesViz.plot(rules, method}\OperatorTok{=}\StringTok{"scatter"}\NormalTok{)}
\NormalTok{image\_png(gg)}
\end{Highlighting}
\end{Shaded}

In Figure 2, we see that high lift rules have typically relative low
support and the process of association rule generation from frequent
itemsets results in characteristic streaks of rules in the
support/confidence space with similar items in the LHS and RHS.

Plots can also be saved as an image using \texttt{ggsave()} (from the R
package \texttt{ggplot2}).

\begin{Shaded}
\begin{Highlighting}[]
\NormalTok{ggsave }\OperatorTok{=}\NormalTok{ packages.importr(}\StringTok{\textquotesingle{}ggplot2\textquotesingle{}}\NormalTok{).ggsave}
\NormalTok{ggsave(gg, }\BuiltInTok{file} \OperatorTok{=} \StringTok{"scatterplot.png"}\NormalTok{)}
\end{Highlighting}
\end{Shaded}

\begin{figure}

{\centering \includegraphics[width=0.6\linewidth]{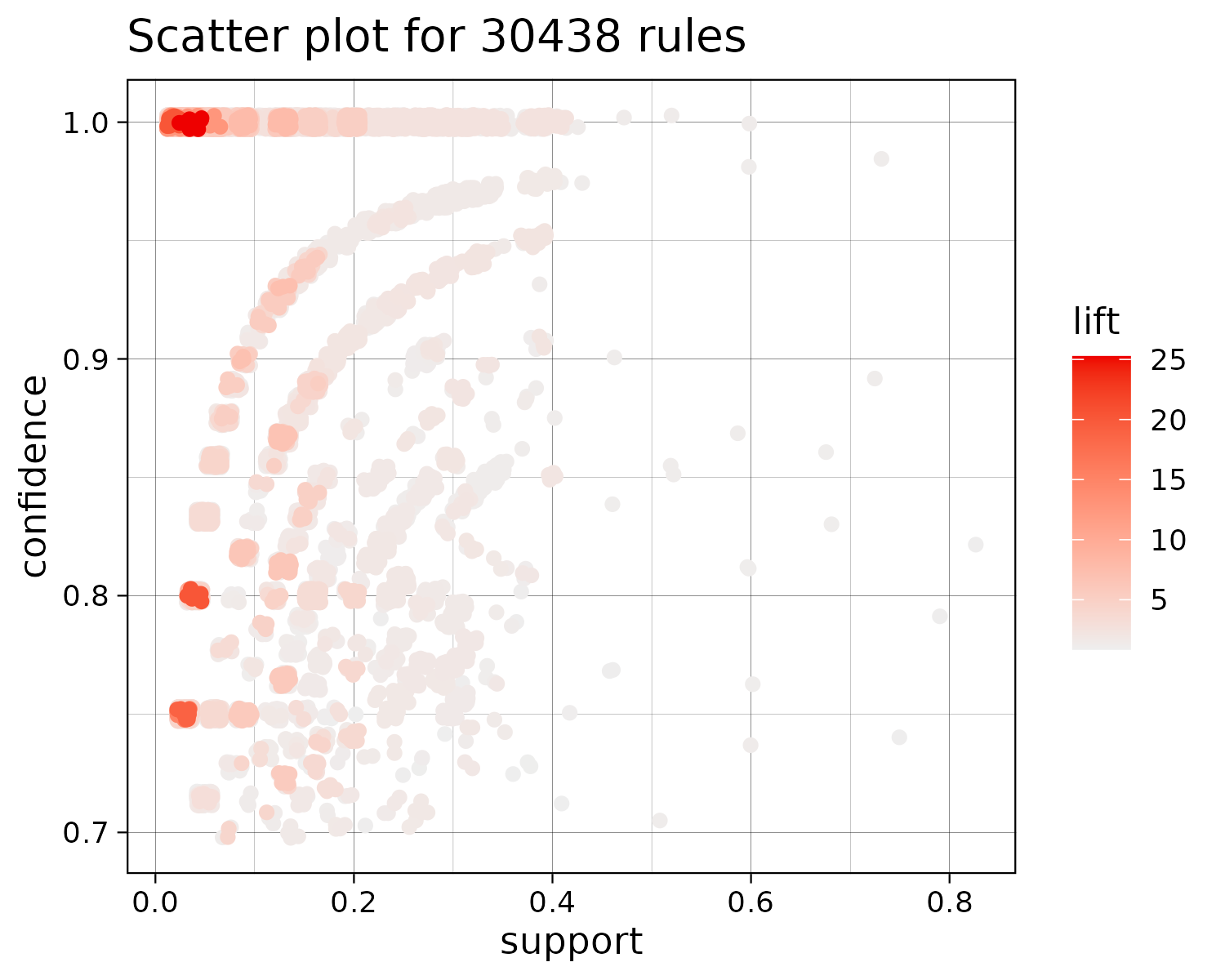} 

}

\caption{Scatter plot of the rules}\label{fig:scatter}
\end{figure}

Another visualization that is appropriate for large rule sets is a
matrix visualization grouped by LHS itemsets introduced by Hahsler and
Karpienko (2016). Figure 3 shows a that the highest-lift rule groups in
the top-lift relate eggs and a different number of legs with the animal
types amphibian and reptile. Lift quickly decreases as we move down,
while support generally increases.

\begin{Shaded}
\begin{Highlighting}[]
\NormalTok{gg }\OperatorTok{=}\NormalTok{ plot(rules, method}\OperatorTok{=}\StringTok{"grouped"}\NormalTok{)}
\NormalTok{image\_png(gg)}
\end{Highlighting}
\end{Shaded}

\begin{figure}

{\centering \includegraphics[width=0.8\linewidth]{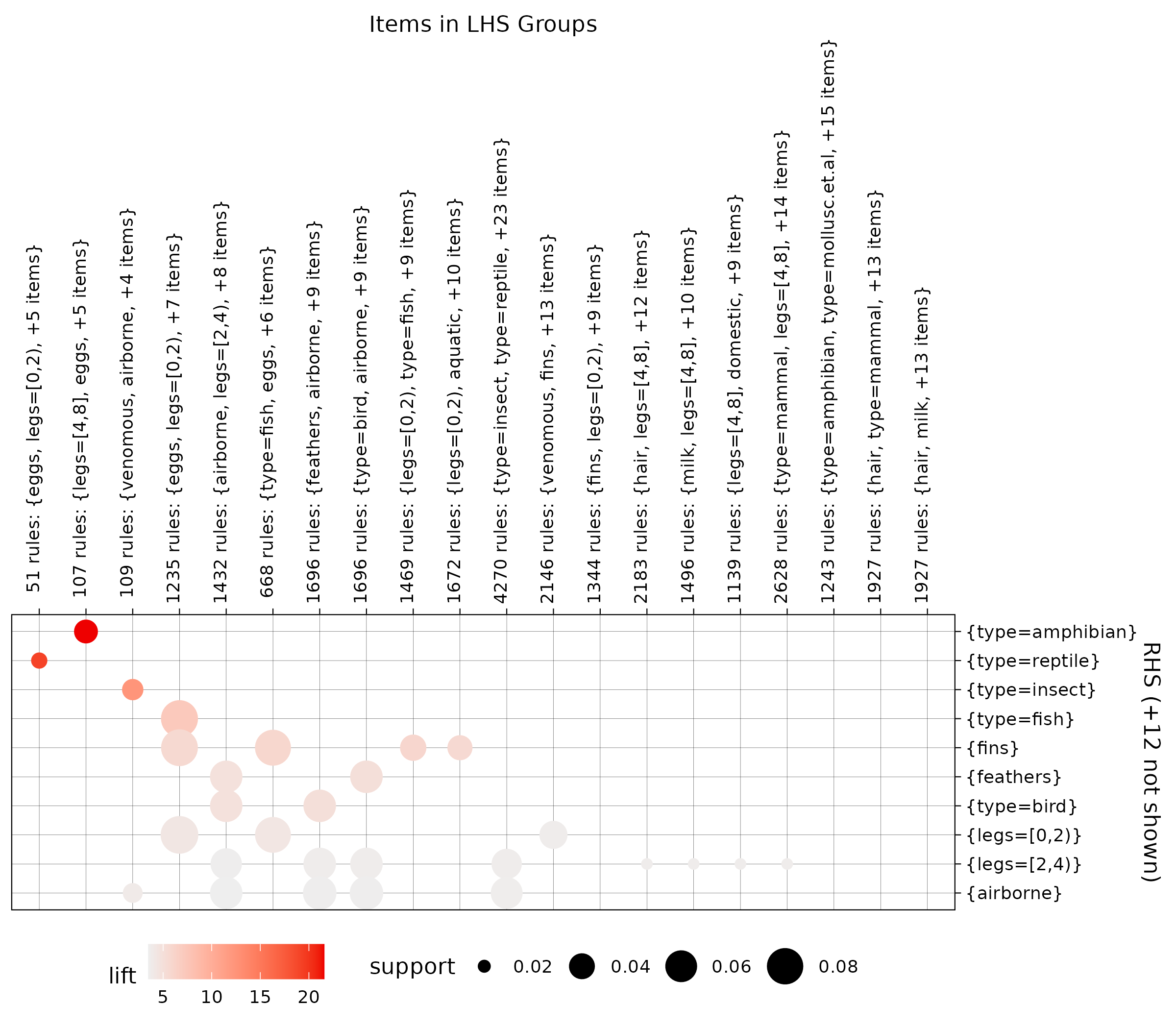} 

}

\caption{Grouped matrix visualization}\label{fig:grouped}
\end{figure}

Another popular visualization of a set of rules is as a graph. This
visualization method is only useful for relatively small rules sets. We
therefore filter the rules first to keep only rules with an animal-type
item in the RHS. This can be again done by list comprehension and
slicing.

\begin{Shaded}
\begin{Highlighting}[]
\NormalTok{type\_rules }\OperatorTok{=}\NormalTok{ rules[[}\StringTok{\textquotesingle{}type\textquotesingle{}} \KeywordTok{in}\NormalTok{ x }\ControlFlowTok{for}\NormalTok{ x }\KeywordTok{in}\NormalTok{ rules.rhs().labels()]]}
\end{Highlighting}
\end{Shaded}

We plot now the top 100 rules by confidence as a graph. Rules are
represented as bubbles with the size proportional to rule support and
the color proportional to lift. In Figure 4, we can identify four groups
of rules with the items for the types bird, mammal, fish and insect in
the RHS. While mammal rules are generally high-support and low-lift,
insects rules have very high lift.

\begin{Shaded}
\begin{Highlighting}[]
\NormalTok{rules\_top\_100 }\OperatorTok{=}\NormalTok{ arules.sort(type\_rules, by }\OperatorTok{=} \StringTok{\textquotesingle{}confidence\textquotesingle{}}\NormalTok{)[}\DecValTok{0}\NormalTok{:}\DecValTok{100}\NormalTok{]}
\NormalTok{gg }\OperatorTok{=}\NormalTok{ arulesViz.plot(rules\_top\_100, method}\OperatorTok{=}\StringTok{"graph"}\NormalTok{)}
\end{Highlighting}
\end{Shaded}

\begin{figure}

{\centering \includegraphics[width=0.7\linewidth]{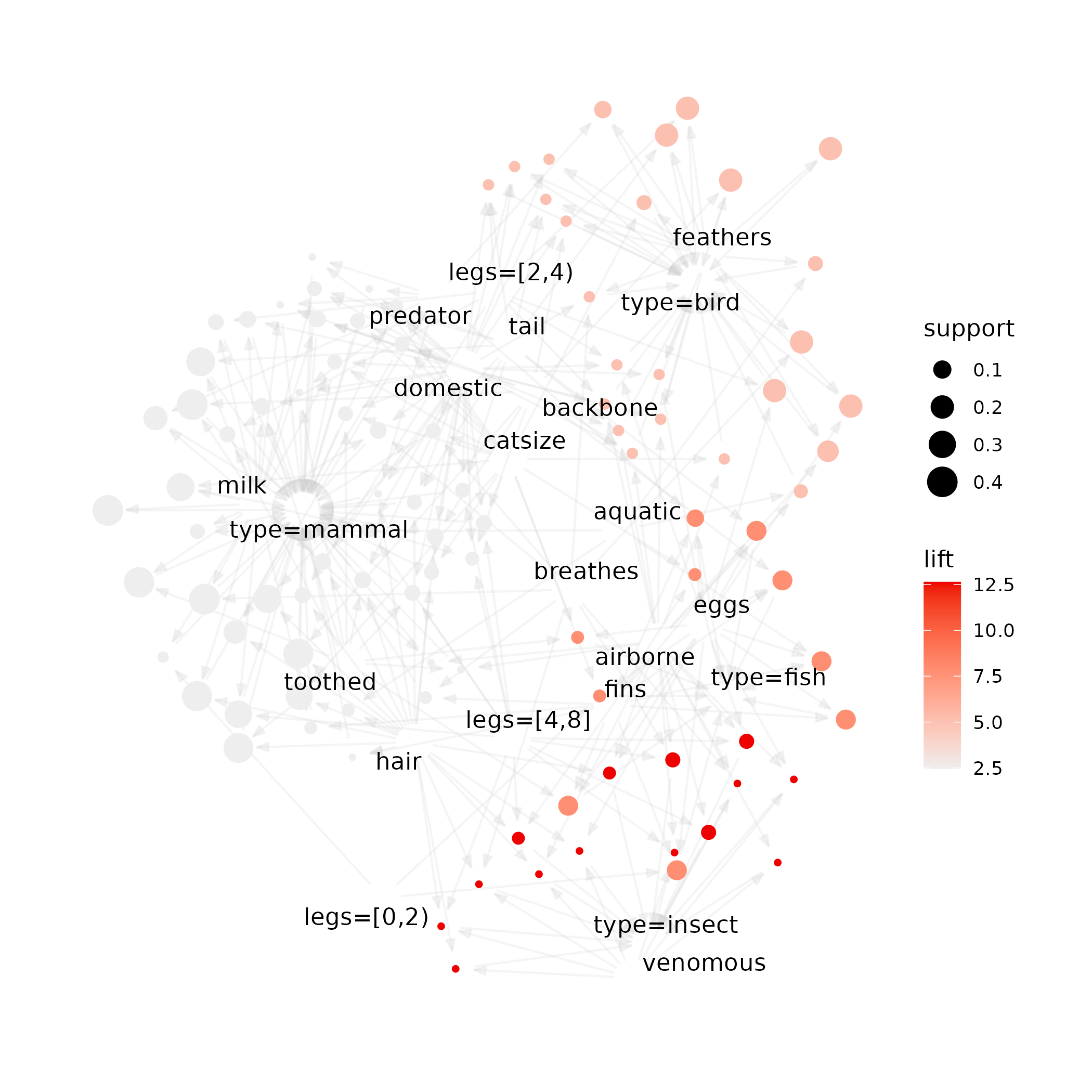} 

}

\caption{Plot rules as a graph}\label{fig:graph}
\end{figure}

\hypertarget{interactive-visualizations}{%
\subsection{Interactive
Visualizations}\label{interactive-visualizations}}

\texttt{arulesViz} supports interactive visualizations (Hahsler 2017)
using interactive HTML widgets (Vaidyanathan et al. 2023), which are
self-contained HTML pages that can be shared as HTML or embedded
directly into Jupyter notebooks using an \texttt{IFrame}.

A very powerful tool to analyze sets of rules is just a sortable table.
\texttt{inspectDF()} produces a widget where rules can be interactively
filtered and sorted. Most static plots available in module
\texttt{arulesViz} can also be created as interactive widgets using
packages like \texttt{plotly} (Sievert 2020) and \texttt{visNetwork}
(Almende B.V. and Contributors and Thieurmel 2022). The interactive
widget created by the following code is available at
\url{https://mhahsler.github.io/arulespy/examples/rules.html}.

\begin{Shaded}
\begin{Highlighting}[]
\ImportTok{from}\NormalTok{ IPython.display }\ImportTok{import}\NormalTok{ IFrame}
\ImportTok{import}\NormalTok{ rpy2.robjects.packages }\ImportTok{as}\NormalTok{ packages}
\NormalTok{saveWidget }\OperatorTok{=}\NormalTok{ packages.importr(}\StringTok{\textquotesingle{}htmlwidgets\textquotesingle{}}\NormalTok{).saveWidget}

\NormalTok{m }\OperatorTok{=}\NormalTok{ inspectDT(rules)}

\NormalTok{saveWidget(m, }\StringTok{"datatable.html"}\NormalTok{, selfcontained }\OperatorTok{=} \VariableTok{True}\NormalTok{)}
\NormalTok{IFrame(}\StringTok{"datatable.html"}\NormalTok{, }\StringTok{"100\%"}\NormalTok{, }\DecValTok{800}\NormalTok{)}
\end{Highlighting}
\end{Shaded}

\hypertarget{low-level-interface}{%
\section{Low-level Interface}\label{low-level-interface}}

The \texttt{arules} module of \texttt{arulepy} exports the rpy2
interface to the \texttt{arules} library using the symbol \texttt{R}.
This provides a complete low-level interface to all arules functions
(see arules reference manual at
\url{https://mhahsler.r-universe.dev/arules\#reference}). Note that the
low-level interface expects all parameters to be R/rpy2 data type and
also returns them. Automatic conversion is not provided, but the helper
function \texttt{a2py()} can be used as a convenient way to convert R
data types into Python data types.

In the following, we create a set of 1000 random transactions and
convert the R/rpy2 transactions object to a Python object.

\begin{Shaded}
\begin{Highlighting}[]
\ImportTok{from}\NormalTok{ arulespy.arules }\ImportTok{import}\NormalTok{ R, r2py}

\NormalTok{trans }\OperatorTok{=}\NormalTok{ a2py(R.random\_transactions(}\DecValTok{10}\NormalTok{, }\DecValTok{1000}\NormalTok{))}
\BuiltInTok{print}\NormalTok{(trans)}
\end{Highlighting}
\end{Shaded}

\begin{verbatim}
## transactions in sparse format with
##  1000 transactions (rows) and
##  10 items (columns)
\end{verbatim}

The low-level interface also lets the user directly access the sparse
representation. It is automatically transformed into a \texttt{scipy}
sparse array.

\begin{Shaded}
\begin{Highlighting}[]
\ImportTok{from}\NormalTok{ scipy.sparse }\ImportTok{import}\NormalTok{ csc\_array}

\NormalTok{trans.items().as\_csc\_array()}
\end{Highlighting}
\end{Shaded}

\begin{verbatim}
## <10x1000 sparse array of type '<class 'numpy.int64'>'
##      with 2959 stored elements in Compressed Sparse Column format>
\end{verbatim}

Finally, the low-level interface also lets the user manually define the
contents of objects in arules. For example, we create a set of three
rules. This is a low-level operation, since it translates item labels in
Python lists into the sparse representation used internally.

\begin{Shaded}
\begin{Highlighting}[]
\NormalTok{lhs }\OperatorTok{=}\NormalTok{ [}
\NormalTok{    [}\StringTok{\textquotesingle{}hair\textquotesingle{}}\NormalTok{, }\StringTok{\textquotesingle{}milk\textquotesingle{}}\NormalTok{, }\StringTok{\textquotesingle{}predator\textquotesingle{}}\NormalTok{],}
\NormalTok{    [}\StringTok{\textquotesingle{}hair\textquotesingle{}}\NormalTok{, }\StringTok{\textquotesingle{}tail\textquotesingle{}}\NormalTok{, }\StringTok{\textquotesingle{}predator\textquotesingle{}}\NormalTok{],}
\NormalTok{    [}\StringTok{\textquotesingle{}fins\textquotesingle{}}\NormalTok{]}
\NormalTok{]}

\NormalTok{rhs }\OperatorTok{=}\NormalTok{ [}
\NormalTok{    [}\StringTok{\textquotesingle{}type=mammal\textquotesingle{}}\NormalTok{],}
\NormalTok{    [}\StringTok{\textquotesingle{}type=mammal\textquotesingle{}}\NormalTok{],}
\NormalTok{    [}\StringTok{\textquotesingle{}type=fish\textquotesingle{}}\NormalTok{]}
\NormalTok{]}
                          
\NormalTok{r }\OperatorTok{=}\NormalTok{ Rules.new(ItemMatrix.from\_list(lhs, itemLabels }\OperatorTok{=}\NormalTok{ trans),}
\NormalTok{              ItemMatrix.from\_list(rhs, itemLabels }\OperatorTok{=}\NormalTok{ trans))}

\NormalTok{r.addQuality(r.interestMeasure([}\StringTok{\textquotesingle{}support\textquotesingle{}}\NormalTok{, }\StringTok{\textquotesingle{}confidence\textquotesingle{}}\NormalTok{, }\StringTok{\textquotesingle{}lift\textquotesingle{}}\NormalTok{], trans))}

\NormalTok{r.as\_df()}
\end{Highlighting}
\end{Shaded}

\begin{longtable}[]{@{}
  >{\raggedleft\arraybackslash}p{(\columnwidth - 10\tabcolsep) * \real{0.0541}}
  >{\raggedright\arraybackslash}p{(\columnwidth - 10\tabcolsep) * \real{0.2973}}
  >{\raggedright\arraybackslash}p{(\columnwidth - 10\tabcolsep) * \real{0.2027}}
  >{\raggedleft\arraybackslash}p{(\columnwidth - 10\tabcolsep) * \real{0.1486}}
  >{\raggedleft\arraybackslash}p{(\columnwidth - 10\tabcolsep) * \real{0.1892}}
  >{\raggedleft\arraybackslash}p{(\columnwidth - 10\tabcolsep) * \real{0.1081}}@{}}
\toprule()
\begin{minipage}[b]{\linewidth}\raggedleft
\end{minipage} & \begin{minipage}[b]{\linewidth}\raggedright
LHS
\end{minipage} & \begin{minipage}[b]{\linewidth}\raggedright
RHS
\end{minipage} & \begin{minipage}[b]{\linewidth}\raggedleft
support
\end{minipage} & \begin{minipage}[b]{\linewidth}\raggedleft
confidence
\end{minipage} & \begin{minipage}[b]{\linewidth}\raggedleft
lift
\end{minipage} \\
\midrule()
\endhead
1 & \{hair,milk,predator\} & \{type=mammal\} & 0.2 & 1 & 2.46 \\
2 & \{hair,predator,tail\} & \{type=mammal\} & 0.16 & 1 & 2.46 \\
3 & \{fins\} & \{type=fish\} & 0.13 & 0.76 & 5.94 \\
\bottomrule()
\end{longtable}

Most functions in arules are accessible using the Python classes and
their methods. Using the low-level interface will only needed
occasionally or when the user wants to implement new functionality that
performs computation on the underlying data.

\hypertarget{conclusion}{%
\section{Conclusion}\label{conclusion}}

This report introduces the usage of \texttt{arulespy}, a new package
that makes the functionality of the R infrastructure to mine and
visualize association rules available to the Python community. After R
is installed, no further R knowledge is necessary to work with the
package and most functions work as expected by Python programmers and
integrate easily with popular Python tools like Jupyter notebooks.

\hypertarget{references}{%
\section*{References}\label{references}}
\addcontentsline{toc}{section}{References}

\hypertarget{refs}{}
\begin{CSLReferences}{1}{0}
\leavevmode\vadjust pre{\hypertarget{ref-arules:Agrawal+Imielinski+Swami:1993}{}}%
Agrawal, Rakesh, Tomasz Imielinski, and Arun Swami. 1993. {``Mining
Association Rules Between Sets of Items in Large Databases.''} In
\emph{Proceedings of the 1993 ACM SIGMOD International Conference on
Management of Data}, 207--16. Washington, D.C., United States: ACM
Press.

\leavevmode\vadjust pre{\hypertarget{ref-arules:Almende:2022}{}}%
Almende B.V. and Contributors, and Benoit Thieurmel. 2022.
\emph{visNetwork: Network Visualization Using 'Vis.js' Library}.
\url{https://CRAN.R-project.org/package=visNetwork}.

\leavevmode\vadjust pre{\hypertarget{ref-arules:Blake+Merz:1998}{}}%
Asuncion, A., and D. J. Newman. 2007. \emph{{UCI} Repository of Machine
Learning Databases}. University of California, Irvine, Deptartment of
Information; Computer Sciences.
\url{http://www.ics.uci.edu/~mlearn/MLRepository.html}.

\leavevmode\vadjust pre{\hypertarget{ref-arules:Bates:2022}{}}%
Bates, Douglas, Martin Maechler, and Mikael Jagan. 2022. \emph{Matrix:
{S}parse and Dense Matrix Classes and Methods}.
\url{https://CRAN.R-project.org/package=Matrix}.

\leavevmode\vadjust pre{\hypertarget{ref-arules:Bayardo+Agrawal+Gunopulos:2000}{}}%
Bayardo, R., R. Agrawal, and D. Gunopulos. 2000. {``Constraint-Based
Rule Mining in Large, Dense Databases.''} \emph{Data Mining and
Knowledge Discovery} 4 (2/3): 217--40.

\leavevmode\vadjust pre{\hypertarget{ref-arules:Borgelt:2003}{}}%
Borgelt, Christian. 2003. {``Efficient Implementations of {Apriori} and
{Eclat}.''} In \emph{FIMI'03: Proceedings of the IEEE ICDM Workshop on
Frequent Itemset Mining Implementations}.

\leavevmode\vadjust pre{\hypertarget{ref-arules:Charles:2022}{}}%
Gautier, Laurent. 2022. {``{RPy2}: {P}ython to {R} Bridge.''}
\emph{GitHub Repository}. \url{https://github.com/rpy2/rpy2}; GitHub.

\leavevmode\vadjust pre{\hypertarget{ref-arules:Geng:2006}{}}%
Geng, Liqiang, and Howard J. Hamilton. 2006. {``Interestingness Measures
for Data Mining: A Survey.''} \emph{ACM Computing Surveys} 38 (3): 9.
\url{https://doi.org/10.1145/1132960.1132963}.

\leavevmode\vadjust pre{\hypertarget{ref-arules:Hahsler:2005}{}}%
Hahsler, Michael. 2005. {``A Probabilistic Comparison of Commonly Used
Interest Measures for Association Rules.''}
\url{https://mhahsler.github.io/arules/docs/measures}.

\leavevmode\vadjust pre{\hypertarget{ref-arules:Hahsler:2017}{}}%
---------. 2017. {``Arules{V}iz: {I}nteractive Visualization of
Association Rules with {R}.''} \emph{R Journal} 9 (2): 163--75.
\url{https://doi.org/10.32614/RJ-2017-047}.

\leavevmode\vadjust pre{\hypertarget{ref-arules:Hahsler:2023}{}}%
---------. 2023. \emph{Python Interface to the {R} Package Arules}.
\url{https://pypi.org/project/arulespy/}.

\leavevmode\vadjust pre{\hypertarget{ref-arules:Hahsler:2011}{}}%
Hahsler, Michael, Sudheer Chelluboina, Kurt Hornik, and Christian
Buchta. 2011. {``The Arules {R}-Package Ecosystem: {A}nalyzing
Interesting Patterns from Large Transaction Datasets.''} \emph{Journal
of Machine Learning Research} 12: 1977--81.
\url{https://jmlr.csail.mit.edu/papers/v12/hahsler11a.html}.

\leavevmode\vadjust pre{\hypertarget{ref-arules:Hahsler+Gruen+Hornik:2005b}{}}%
Hahsler, Michael, Bettina Grün, and Kurt Hornik. 2005. {``Arules -- {A}
Computational Environment for Mining Association Rules and Frequent Item
Sets.''} \emph{Journal of Statistical Software} 14 (15): 1--25.
\url{http://www.jstatsoft.org/v14/i15/}.

\leavevmode\vadjust pre{\hypertarget{ref-arules:Hahsler2016c}{}}%
Hahsler, Michael, and Radoslaw Karpienko. 2016. {``Visualizing
Association Rules in Hierarchical Groups.''} \emph{Journal of Business
Economics} 87 (3): 317--35.

\leavevmode\vadjust pre{\hypertarget{ref-arules:Hofmann+Wilhelm:2001}{}}%
Hofmann, Heike, and Adalbert Wilhelm. 2001. {``Visual Comparison of
Association Rules.''} \emph{Computational Statistics} 16 (3): 399--415.

\leavevmode\vadjust pre{\hypertarget{ref-arules:Lenca:2007}{}}%
Lenca, Philippe, Benoît Vaillant, Patrick Meyer, and Stephane Lallich.
2007. {``Association Rule Interestingness Measures: Experimental and
Theoretical Studies.''} In \emph{Quality Measures in Data Mining},
edited by Fabrice J. Guillet and Howard J. Hamilton, 51--76. Berlin,
Heidelberg: Springer Berlin Heidelberg.
\url{https://doi.org/10.1007/978-3-540-44918-8_3}.

\leavevmode\vadjust pre{\hypertarget{ref-arules:R:2005}{}}%
R Development Core Team. 2005. \emph{R: A Language and Environment for
Statistical Computing}. Vienna, Austria: R Foundation for Statistical
Computing. \url{http://www.R-project.org}.

\leavevmode\vadjust pre{\hypertarget{ref-raschkas_2018_mlxtend}{}}%
Raschka, Sebastian. 2018. {``MLxtend: Providing Machine Learning and
Data Science Utilities and Extensions to Python's Scientific Computing
Stack.''} \emph{The Journal of Open Source Software} 3 (24).
\url{https://doi.org/10.21105/joss.00638}.

\leavevmode\vadjust pre{\hypertarget{ref-arules:Sievert:2020}{}}%
Sievert, Carson. 2020. \emph{Interactive Web-Based Data Visualization
with r, Plotly, and Shiny}. Chapman; Hall/CRC.
\url{https://plotly-r.com}.

\leavevmode\vadjust pre{\hypertarget{ref-arules:Tan:2004}{}}%
Tan, Pang-Ning, Vipin Kumar, and Jaideep Srivastava. 2004. {``Selecting
the Right Objective Measure for Association Analysis.''}
\emph{Information Systems} 29 (4): 293--313.
\url{https://doi.org/10.1016/s0306-4379(03)00072-3}.

\leavevmode\vadjust pre{\hypertarget{ref-arules:Vaidyanathan:2023}{}}%
Vaidyanathan, Ramnath, Yihui Xie, JJ Allaire, Joe Cheng, Carson Sievert,
and Kenton Russell. 2023. \emph{Htmlwidgets: {HTML} Widgets for {R}}.
\url{https://CRAN.R-project.org/package=htmlwidgets}.

\leavevmode\vadjust pre{\hypertarget{ref-arules:Wickham:2016}{}}%
Wickham, Hadley. 2016. \emph{Ggplot2: {E}legant Graphics for Data
Analysis}. Springer-Verlag New York.
\url{https://ggplot2.tidyverse.org}.

\end{CSLReferences}

\bibliographystyle{unsrt}
\bibliography{paper.bib}

\end{document}